\def\entitle#1#2{{In their paper entitled {\it #1} #2}}
\def\app#1{\noindent{\bf #1}\par\smallskip\noindent}
\def\et{{\it et al.}}

\def\kpc{$h^{-1}${\rm kpc}}
\def\ha{hereafter~}

\font\small=cmr8
\font\smalll=cmr6

\def\R{{${\cal R}$}}
\def\Ho{${\rm H}_0$}
\def\H0#1{{H$_0$ = #1 km~s$^{-1}$ Mpc$^{-1}$}}

\def\CIV{{\rm C{\small IV}}}

\def\HII{{\rm H{\small II}}}
\def\MgII{{\rm Mg{\small II}}}
\def\OII{{\rm [O{\small II}]}}
\def\TABLE{T{\small ABLE\ }}
\def\zgal{{${\rm z}_{\rm gal}$}}
\def\zabs{{${\rm z}_{\rm abs}$}}
\def\zform{{$z_{\rm form}$}}
\def\Lemaitre{Lema\^\i tre}
\def\Boisse{Boiss\'e}
\def\fig#1#2{{Figure~#1~#2{\small )}}}
\def\L#1#2{#2)}
\def\rr#1{{#1}}
\vglue 0.2truein

\input epsf
\MAINTITLE={Optically identified QSO absorption systems and galaxy evolution}
\AUTHOR={Ulrich Lindner, Uta Fritze -- von Alvensleben, Klaus J. Fricke}
\INSTITUTE{Universit\"atssternwarte, Geismarlandstra\ss e 11, 
              D--37083 G\"ottingen, Germany}  
\ABSTRACT
{A comprehensive set of all available magnitudes, colors and
redshifts for optically identified QSO absorbers is compiled
from the literature. This results in a largely unbiased sample
of galaxies at high redshifts, the size of the sample being
reasonably large to allow for a first comparison with galaxy
evolution models. We use our evolutionary spectral synthesis 
models describing various spectral types of galaxies with
appropriate star formation histories.

For a standard cosmology \H0{50}, $\Omega_0 = 1$, we find
that all identified QSO absorbers up to a redshift $\simeq 2$
do fall within the evolving range of models which had been
shown earlier to give good agreement with nearby galaxies
of various spectral types.

The most striking result is that {\bf all spectral types}
of galaxies from E through Sd seem to be present in this 
absorber candidate sample with the bulk of the objects being 
spirals of type Sa to Sc.

We also show that such kind of high redshift galaxy sample
should allow for significant restrictions of the cosmological
parameters. While models with \H0{50}, $\Omega_0 = 1$ and 
any redshift of galaxy formation \zform\ $= 5 \ldots 10$
give reasonable agreement with observed luminosities and
colors, present data seem to exclude the combination
\H0{50}, $\Omega_0 = 0.1$. A low $\Omega_0$ together
with a value of $H_0 > 50$ might still be viable. 
Additional data, and in particular
observations of absorbing galaxies in more than one
passband, are required for further constraints. 
\rr{We realize that the present results depend on the 
adopted evolutionary models and on the input physics used
which might be better constrained by future observations.} }

\KEYWORDS={Galaxies: evolution -- Quasars: absorption lines -- absorption 
line systems: Optical identification}   
\THESAURUS={ 11(11.05.2, 11.17.1)}      
\OFFPRINTS={Ulrich Lindner, Uni\-ver\-si\-t\"ats\-stern\-warte, 
Geismarlandstra\ss e 11, D--37083 G\"ottingen, Germany,
e-mail: ulindner\at uni-sw.gwdg.de}             
\DATE={Received 17 August 1995 / Accepted 16 April 1996} 
\maketitle
\MAINTITLERUNNINGHEAD{Optically identified QSO absorber systems 
                      and galaxy evolution}
\AUTHORRUNNINGHEAD{U. Lindner \et}

\titlea{Introduction}
Among others, absorption lines from the doublets of \MgII\ and \CIV, 
damped Lyman $\alpha$ lines (\ha DLA), and Lyman Limit Systems 
(\ha LLS) are seen in the spectra of quasars.
By now, interpretations seem to converge to the view
that the \MgII\ and \CIV\ doublets arise in extended gaseous
halos of intervening galaxies, while DLA's are formed in 
(proto --) galactic disks (Steidel, 1993). 

This picture remained hypothetical until the QSO absorbers, 
i.e. the galaxies causing the absorption lines were actually 
detected. Pioneering work in the optical was done by 
Bergeron \& \Boisse\ (1991). By now, there are more than a
dozen of publications reporting on optical identifications 
of QSO absorber systems. Their number increases rapidly. 
We have compiled all relevant data available to us. 
Information about the data we use is given in \S 3
and the sources from which our data are drawn are 
briefly reported in the 
appendix\fonote{The appendix is also available in
electronic form at the CDS via anonymous ftp 
to cdsarc.u-strasbg.fr (130.79.128.5) or via
http://cdsweb.u-strasbg.fr/Abstract .html}.

A sample of optically identified QSO absorbers 
has, at least, two advantages over others, e.g. magnitude limited 
samples, (i) it is scarcely biased to selection effects with 
respect to luminosity, surface brightness, radio emission, 
etc. (cf. Steidel \et\ 1993, Steidel 1995)
and (ii) it gives access to galaxies with high redshifts.

This paper is a first attempt to compare a comprehensive
set of magnitudes and colors of optically identified QSO
absorbers and absorber candidates with galaxy evolution
models. Models are briefly described in \S 2.
The aim of this study is first to investigate the properties
of this QSO absorbing galaxy sample (cf. \S 4) and, second,
to show that this kind of high redshift galaxy sample should
allow for serious restrictions of cosmological
parameters (\S 4.3).
We discuss some problems in \S 5 and outline future
prospects in \S 6. \S 7 summarizes our conclusions.
 
\titlea{Models of galaxy evolution}
Here, we only give a brief outline of our galaxy evolution 
models, which have been described in detail earlier by 
Fritze -- von Alvensleben (1989), Kr\"uger \et\ (1991) and
Fritze -- von Alvensleben \& Gerhard (1994).

\titleb{General description of models}
Starting from an initial gas cloud stars are formed continuously
in a 1--zone model according to a given star formation law.
The distribution of the total astrated mass to discrete stellar masses
in the range 0.04 M$_\odot$ ... 85 M$_\odot$ is described by a
Scalo (1986) IMF. 

Using recent stellar evolutionary tracks for solar metallicity
from the Geneva group (Maeder, 1991), the population of the HRD is
calculated as a function of time. 
Synthetic galaxy spectra are obtained 
by assigning stellar spectral types and luminosity classes
to some 40 cells in the HRD and by superimposing observed
stellar spectra from Gunn \& Stryker's (1983) library weighted
by the number of stars present in each cell at a given time.
Convolution of synthetic galaxy spectra with filter functions 
yields absolute magnitudes and colors of the model galaxies.

Basic parameters of this kind of models are the initial mass 
function (\ha IMF) and the star formation (\ha SF) laws. For 
galaxies of various spectral types, we use different parametrisations 
of their SF histories following Sandage (1986).
While an elliptical galaxy is described by a SF rate declining
exponentially in time with a characteristic time scale of 1 Gyr, 
the SF rate in spiral galaxies is a linear function of the gas
content with characteristic time scales ranging from 
1, 2, 3, 10 to 16 Gyr for S0, Sa, Sb, Sc and Sd, respectively. 
Gas recycling due to stellar winds, supernovae and 
planetary nebula is consistently taken into account.

These SF laws together with a Scalo IMF have been shown to give
good agreement with observations of 
the respective galaxy types in the Virgo cluster and in the field
in terms of broad band colors UBVRI, gaseous emission lines,
M/L ratios, gas content, etc. (Fritze -- von Alvensleben 1989, 
Kr\"uger \et\ 1991 \& 1995,
Fritze -- von Alvensleben \& Gerhard (1994).
Furthermore, we find detailed agreement with Kennicutt's (1992)
template galaxy spectra for the various Hubble types.

\rr{For comparable characteristic time scales in the
SF laws, the time evolution of our model galaxies
agrees to within 0.1 mag with Bruzual \& Charlot's (1993)
models which are based on slightly different stellar evolution tracks.
Except for very early phases ($z \ge 1$) we obtain somewhat 
bluer colors at the largest wavelengths.}  

It should be stressed that in our context the different {\bf Hubble
types} of galaxies as described by their respective SF laws are
meant to be {\bf spectral types} rather than morphological types.
Only for nearby galaxies the correspondence between morphological
types and spectroscopic types has been shown, it remains to
be investigated for high redshift galaxies.
We have also studied the chemical evolution of individual
elements in the interstellar medium (\ha ISM) for these same 
SF laws and found satisfactory agreement with observations of
\MgII-- and \CIV--systems (Fritze -- von Alvensleben \et\ 1989 \& 1991)
and with abundances derived for DLA systems as well
(Fritze -- von Alvensleben 1994) over a redshift range $0 \le z \le 4$
for a Scalo--IMF.

\rr{It is clear, however, that all the results presented below
depend quantitatively on our specific models. Should future
observational evidence impose changes, e.g. in the IMF, the SF
laws, or in the stellar input tracks, our conclusions may have
to be modified appropriately.}

To transform time dependent absolute magnitudes into
observable apparent magnitudes $m_\lambda(z)$ as a function 
of redshift a cosmological model has to be specified. 
Here, we adopt a Friedmann--\Lemaitre\ model with vanishing 
cosmological constant ($\Lambda_0 = 0$), so \Ho\ and 
$\Omega_0$ ($=2q_0$ in the case of $\Lambda_0 = 0$)
are the cosmological parameters together with the redshift
\zform\ where star formation in our model 
galaxies is assumed to start. 

The relation between redshift and time is given by the
standard formula for the {\it Hubble--time}
$T_H ( z, H_0, \Omega_0 )$. Then, for given $(H_0, \Omega_0 )$ the age of a
galaxy formed at redshift \zform\ is calculated from 

\smallskip\noindent \qquad  $t_{gal}(z) = T_H (z) - T_H $(\zform).

\smallskip\noindent 
The relation between redshift and distance is 
given by the standard formula for the {\it luminosity distance}
$D_L ( z, H_0, \Omega_0 )$  which is used to
calculate the {\it bolometric distance modulus} 

\smallskip\noindent \qquad  $BDM(z)= 5 ~\log( D_L ) ~+~ 25$.

\smallskip\noindent  
Finally, the observable apparent magnitude at a
wavelength (filter band) $\lambda$ writes
\medskip\noindent\qquad $m_{\lambda}(z)=BDM(z)+k_\lambda(z)+e_\lambda(z)+M_{\lambda}(0,t_0)$
\medskip\noindent
with $t_0=t_{gal}(z=0)$. The evolutionary correction 

\smallskip\noindent
\qquad $e_\lambda(z):=M_{\lambda}(z,t_{gal}(z) )-M_{\lambda}(z,t_0)$ 

\smallskip\noindent
takes into account the variation
of the galaxy spectrum with time. The cosmological correction

\smallskip\noindent
\qquad $k_\lambda(z):=M_{\lambda}(z,t_0)-M_{\lambda}(0,t_0)$

\smallskip\noindent 
accounts for the influence of the expansion of the Universe.

It is important to stress that both
the evolutionary and cosmological corrections do not only
depend on the cosmological parameters but also on the spectral 
type of the galaxies (cf. also Rocca--Volmerange \& Guiderdoni, 1988).
Before red-shifting our model galaxies, we
calibrate their absolute magnitudes M$_B$ to mean luminosities
of galaxies in the Virgo cluster as determined by 
Sandage \et\ (1985a, 1985b) for the various Hubble types.

\titleb{Implications from models}

For this paper, apparent magnitudes as a function of redshift are 
calculated from our models for two different photometric systems.
Results of our models for Johnson R and  Thuan \& Gunn r are 
presented in Figure~1. 
Late type galaxies (Sd) not only are significantly fainter
than mean early type galaxies (E) at low redshift, but they
get fainter with increasing redshift more rapidly than those.
This is the reason why in magnitude limited optical samples late type
galaxies with redshifts $z \ge 0.5$ are very rare.

\begfig 5.0cm
\vskip -7.5cm
\epsfysize=9.5cm
\hskip 0.1cm
{\epsffile{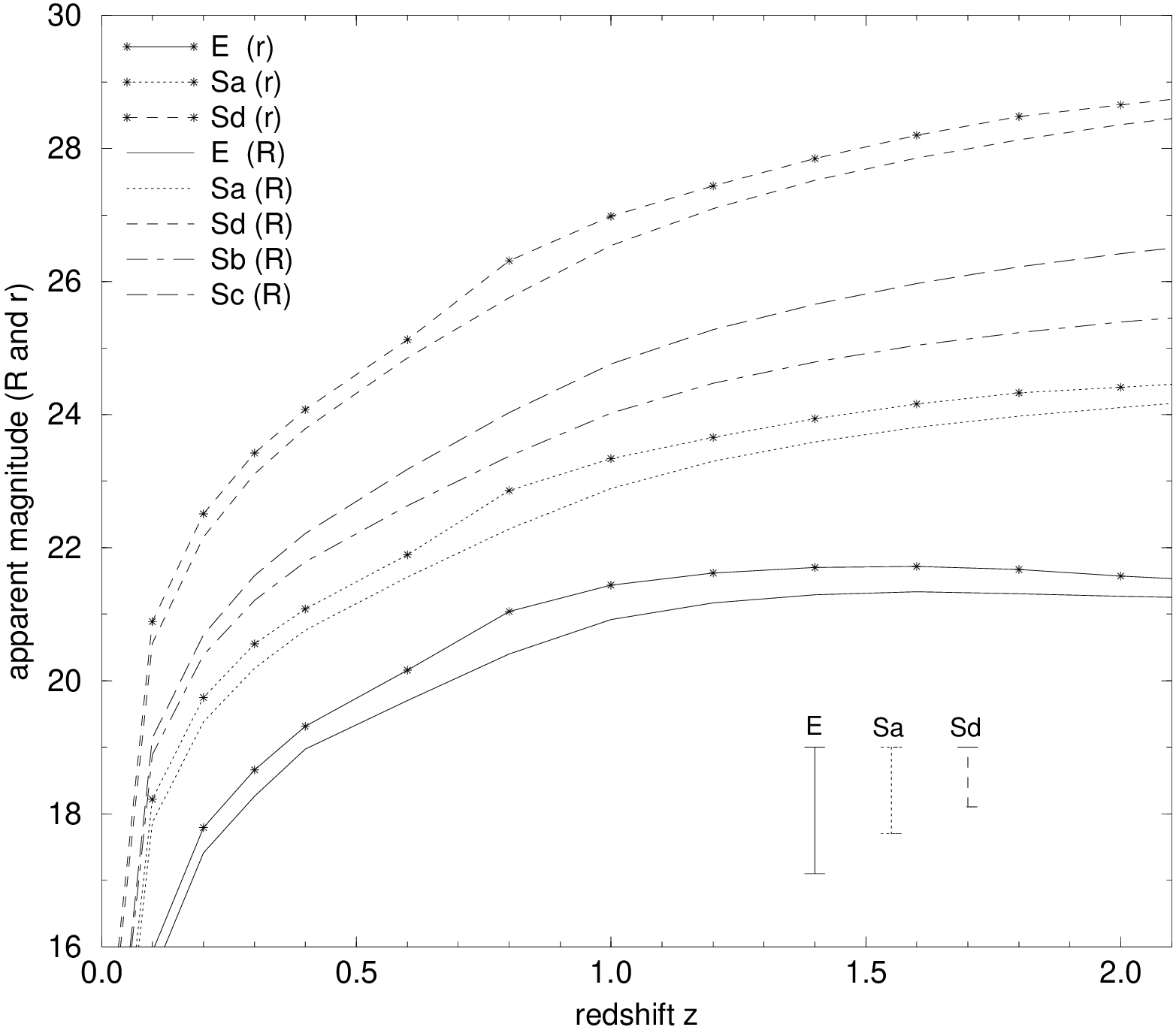}}
\vskip 0.3cm 
\figure{1}
{Results from galaxy evolution models with cosmological 
parameters \H0{50} and $\Omega_0 = 1$ using two different
filter response functions: Johnson R and Thuan \& Gunn r. 
Thuan \& Gunn r(z) curves are marked with small filled circles. 
The luminosity range of the various galaxy types as derived
from their luminosity functions is indicated in the lower right corner} 
\endfig

The redshift evolution of apparent magnitudes for E and Sd
models confine the range of normal galaxies from the upper
and lower luminosity limit. 
In an analogous way, the color evolution with redshift of our
E and Sd models bracket the color range of normal galaxies from
the red and blue, respectively. Observational data are expected
to somewhere fall within the region between our
E and Sd models. In the lower right corner of Figure~1 we
indicate the luminosity range of $\pm \sigma_R$ for galaxies
of various Hubble types as derived from Sandage's (1985a\&b) 
luminosity functions.

Figure~1 also shows that over the redshift range from 0.2 to 2
the difference between Johnson's R and Thuan \& Gunn's r is
approximately constant, $r - R \approx  0.4$, independent of
the spectral type.

We find considerable differences in these
results for various cosmological parameters $H_0$ and $\Omega_0$.
This is shown in Figure~4 and discussed in \S 4.3. 

We adopt a redshift of galaxy formation \zform\ $= 5$ 
throughout this investigation. Smaller values seem to
contradict observations of very high redshift galaxies
(predominantly radio galaxies), and for a larger \zform\ 
the differences in evolution time for the redshift range 
$0.2 < z < 2$ are very small and thus do not influence
the results considerably. 
Previous work (Guiderdoni, 1986 and
Fritze v. -- Alvensleben, 1989) also shows
that differences in the results
using \zform\ $= 5$ and \zform\ $= 10$ 
are negligible in the redshift range 
$0.2 < z < 2$.

\titlea{Data about optically identified QSO absorbers}

We have compiled from the literature all available data on 
observations of optically identified QSO absorption
systems. Data have been collected from 17 different 
publications, details which are important for our purposes
are summarized in the appendix (electronically published). 
A brief overview with the
most relevant data for our current investigation is
given in Table~1 and is detailed in \S 3.1.

These data from various authors are very inhomogeneous
making the comparison between various data sets as well
with results from model calculation rather difficult.

In this work we confine ourselves to the interpretation
of magnitudes and colors of the high redshift galaxy 
sample detected in search for QSO absorbers. A detailed
investigation of individual absorbers, also combining
chemical information (e.g. column densities, ionisation 
state, etc.) derived from the absorber imprint on the QSO
spectrum with spectrophotometric information obtained
for the optically identified absorber will be the 
subject of a forthcoming paper.

In the following subsections data on apparent magnitudes 
and redshifts for absorbing galaxies observed by
various authors are discussed in more detail.

\begtabfullwid \tabcap{1}
{Published observations of optically identified QSO 
absorbers (cf. \S 3.1.): 
Number of QSO fields being investigated, 
number of absorber systems (column ``\zabs''), 
number of spectroscopically measured galaxy redshifts (\zgal) 
and apparent magnitudes (Thuan/Gunn, Johnson or others) 
in the literature}
\halign to \hsize {\hfil#\ &#\hfil&#\hfil&\hfil#&
\hfil#&\hfil#&\hfil#&\hfil#&\hfil#&\hfil#&\hfil#&\hfil#& \quad #\hfil\cr
\noalign {\medskip}
\noalign{\hrule}
\noalign{\medskip}
 id. no.  & author   & absorber & QSOs & \ \zabs\ & \ \zgal\ & g & r & i & V & R & K & filter \cr
 (1) & (2) & (3) & (4) & (5) & (6) & \quad (7) & \ (8) & \ (9) & \ (10) & \ (11)& \ (12) & \ (13) \cr
\noalign {\medskip}
\noalign{\hrule}
\noalign{\medskip} 
\L{225}{1} & Lanzetta \& Bowen, 1990 &  \MgII &  13 &  16  & 0  &  0 &  $^{1)}$14 &  0 &  0 &  0 &  0 & r band, $m_r$\cr
\L{192}{2} & Yanny \et,   1990  & \MgII &  8  &  8  &  8  &  0 &  0 &  0 &  0 & $^{2)}$37 &  0 & R, div. \cr
\L{184}{3} & Bergeron \& \Boisse, 1991 & \MgII & 19 & 21 & 16 &  0 & \bf{21} &  0  & \bf{9} &  0 &  0 & $m(r)$,$m_r$, $m_V$ \cr
\L{191}{4} & Bechtold \& Ellingson, 1992  & \MgII & 10 & 2 &  34 &  0 & 34  &  0  &  0  &  0 &  0 & r, $m_r$ \cr
\L{185}{5} & Bergeron \et, 1992     & \MgII &  2 &  4 &  5 &  0 & \bf{6} & \bf{6} & 0 & 0 & 0 & r band, r, Gi \cr
\noalign{\medskip} 
\L{224}{6} & Nelson \& Malkan, 1992 & \MgII  &  18 &  19  &  6 &  43 &  0 &  0 &  0 &  0 &  0 & Gunn g \cr
\L{190}{7} & Steidel \& Dickinson, 1992 & \MgII & 1 & 3 & 6 & \bf{6} & 0 & \bf{6} &  0 & \bf{6} &  0 & g, \R, i , (AB) \cr
\L{223}{8} & Steidel \& Hamilton, 1992 & div.$^{3)}$ & 1 &  2  &  0  &  2 &  0 &  0 & 0 &  2 &  0 & $U_n$, G, \R, (AB)  \cr
\L{226}{9} & Yanny \& York, 1992    & \MgII  &  3   &  3  &  0  &  0 &  0 &  0 & \bf{3} & \bf{7} & 0 & R \cr 
\L{165}{10} & Drinkwater \et, 1993   & \MgII & 21 & 24 &  0 &  0 &  0    &  0  &  0   & 25 &  0 & R \cr
\noalign{\medskip}
\L{193}{11} & Le Brun \et, 1993    & \MgII & 12 & 17 &  6 &  0    & 24  &  0  &  0 &  0  &  0 & r band, $m_r$ \cr
\L{166}{12} & Spinrad \et, 1993    & div.$^{3)}$ & 1 & 9 & 8 & \bf{7} & \bf{9} &  0  &  0 & 0  &  0 & Gunn g, r\cr
\L{194}{13} & Steidel \et, 1993    & ---$^{4)}$  & 1 & 1 & 2 & 0  & 0 & \bf{2} &  0 & \bf{2} &  0 & \R, i , (AB) \cr
\L{167}{14} & Aragon-Salamanca \et, 1994 \ & \CIV  & 10 & $^{5)}$10 & 0 &  0 &  0  &  0  &  0  &  \bf{9} & \bf{19} & R, K \cr 
\L{221}{15} & Ellingson \et, 1994  & \CIV  &  10  &  4  &  1  &  0 &  7 &  0 &  0 &  0 &  0 & Gunn r \cr
\noalign{\medskip}
\L{175}{16} & Steidel \et, 1994a   & DLA   & 1 & 1 &  0 &  0  &  0  &  0  &  0 & \bf{1} & \bf{1} & \R, K, (AB)\cr
\L{170}{17} & Steidel \et, 1994b   & DLA   & 2 & 3 &  0  &  0 &  0 &  \bf{3} &  0 &  0 &  \bf{2} & $I_{\rm AB}, K_S$\cr
\noalign {\medskip}
\noalign{\medskip}
\noalign{\hrule}
\noalign {\medskip}
\noalign {\small\noindent Notes: 
$^{1)}$ \zabs\ $>$ 14: For some absorption systems no galaxy was found.
$^{2)}$ For 14 from the 37 objects only lower limits are given. Confirming
spectroscopy for 8 galaxy redshifts is taken from Yanny (1990). 
$^{3)}$ Diverse absorber systems (Ly$\alpha$, \CIV\ and \MgII) have been observed.  
$^{4)}$ An absorption system will be searched for the dwarf 
galaxy (~cf. appendix 13{\smalll )}~).
$^{5)}$ Mean value from multiple absorption systems was taken 
(~cf. appendix 14{\smalll )}~).
}           
}
\endtab

\titleb{Overview}
Table~1 gives a brief overview on published 
observations of optically identified QSO absorber systems. 
Each record in the Table is labeled with a number we 
refer to in the text and Figures (column 1). 
In column (2) we give the author's name and in 
column (3) the type of absorption system is indicated. 
The records in the Table are in chronological order. 
In the next three columns, we list the numbers of observed 
QSO fields (4), of detected absorption systems (5), 
and of spectroscopically measured galaxy 
redshifts (6). In the remaining columns the numbers of 
apparent magnitudes determined using Thuan \& Gunn 
(1976) filters g (7), r (8), i (9) and Johnson (1966) filters V (10), 
R (11), K (12), are given. The last column (13) contains some
remarks on the specific passbands (filters) used by the authors.

Note that some objects are counted more than once in
Table~1, i.e. they may be cited by various authors or else, a
specific galaxy can be a possible absorber candidate
to more than one absorption system.
This means that the number of physically different objects is
somewhat smaller than the total number of objects in Table~1.

\titleb{Redshift data}
In column (5) of Table~1 we list the number of 
absorption systems in the respective QSO spectrum and in column 
(6) the number of spectroscopically measured galaxy redshifts
is given. 
Usually more than one absorption line system is known for a single 
QSO spectrum. Thus in many cases the number of absorption 
redshifts is larger than the number of QSO fields being investigated. 

Zero values in column (6) (7 occurrences in Table~1) indicate 
investigations without spectroscopically measured redshifts. In these 
cases statistical methods are used to identify the most probable 
absorber candidate from the image of the QSO environment. Often 
more than one galaxy are possible candidates for
a given absorption line redshift \zabs. 

Less galaxy redshifts than absorption systems reported in 
Table~1 means that spectroscopy was not done for all 
possible absorber galaxies or 
that no galaxy was found at the appropriate redshift. If the number 
of galaxy redshifts (column ``\zgal'') is larger than that 
of absorber redshifts (column ``\zabs'') more than one absorber 
candidate was investigated spectroscopically. In this case some 
galaxies with \zgal\ $\neq$ \zabs\ are in the sample which are not
absorbers. Otherwise, i.e. if \zgal\ $ = $ \zabs\ 
for more galaxies than absorption line systems, two ore more galaxies 
could cause the absorption indicating a small group of galaxies or 
remote parts of a cluster in the line of sight to the QSO.

Spectroscopic redshifts are of course the most reliable data, but 
such measurements are not always available and then
redshifts are inferred from the assumption
that the galaxy in the field closest to the QSO gives rise to the 
absorption line system observed at \zabs\ and thus the
redshift is as uncertain as the absorber identification. 
Therefore we use two different ``reliability classes''
for all $m(z)$ data distinguishing between 
galaxies with {\it spectroscopically 
confirmed redshifts} ({\bf filled symbols} in Figures~2 -- 4) 
and {\it possible candidates} without spectroscopic redshift 
confirmation ({\bf open symbols} in Figures~2 -- 4).

\titleb{Apparent magnitudes}
Table~1 shows that the set of apparent magnitudes 
measured for optically identified absorbing galaxies is extremely 
inhomogeneous. This is a crucial problem for the comparison of
Johnson R and Thuan \& Gunn r magnitudes are
most widely used (altogether 204 occurrences in Table~1),
measurements in other passbands are rarer 
(56 g--, 16 i--, 9 V--, 22 K--magnitudes).

Even more scarce is information about colors, i.e. measurements 
in more than one photometric band. 
9 ($V - r$), 6 ($r -Gi$), 6 ${(g-i)}_{\rm AB}$, 6 ($g-$\R$_{\rm AB}$), 
7 (\R-i)$_{\rm AB}$, 7 ($g - r$), 10 ($R - K$) and 2 ($I_{\rm AB} - K_s$) 
colors are given or can be derived from available magnitudes. 
Observations in more than one passband are 
emphasized by bold numerals 
in Table~1. Some authors give one apparent magnitude and
one color, then the second magnitude is derived from the color.

As shown in Figure~1 and discussed in \S 2.2. we may safely
adopt the relation $r - R \approx 0.4$ to transform all
r band observations to Johnson's R for the purpose of
comparison with our galaxy evolution models.
However, relations of this kind cannot be generalized because 
they depend not only on the shape of the filter response function 
but also on the specific galaxy spectrum. 
Transformations from one photometric system into another as e.g.
given in Landolt--B\"ornstein VI/2b (1982) cannot
be used at redshifts $\ne 0$ for the same reason. 

More troubling than the great variety of filter systems is
the fact that some authors do not give detailed 
information on the photometric system they use. 
We adopt Gunn r if ``r band'', ``r'', ``$m_r$'' or ``m(r)'' 
is mentioned in the text (cf. also column (13) in Table~1)
and Johnson R in case of ``R'' or ``$m_R$''.

Some investigators give maximum transmission wavelength 
$\lambda_0$ and FWHM $\Delta\lambda$ of the filter they use.
For instance, Steidel \et\ throughout use a red filter named \R\ 
characterized by $\lambda = 6930$ and $\Delta\lambda = 1500$ 
(e.g. Steidel \& Dickinson, 1992).
Steidel \& Hamilton (1993) plot the filter response
function of their ($U_n$, G, \R) photometric system and
argue that \R\ is close to the original Johnson R. Thus we
adopt the rough approximation \R\ $ \simeq $ R to include
Steidel \et's data in our compilation.

With the appropriate filter functions our synthetic
galaxy spectra can easily be convolved to give magnitudes in 
any photometric system.
This will be done in a forthcoming paper
investigating individual objects in more detail.

Some authors (cf. Table~1) use AB magnitudes as established
by Oke \& Gunn (1983). According to Lilly \et\ (1991) who give
approximate transformations to Johnson's system for B, V and I 
($B \sim B_{\rm AB} + 0.17$, $V \sim V_{\rm AB}$ and
$I \sim I_{\rm AB} - 0.48$) we derive 
$R \sim R_{\rm AB} - 0.25$ using the F--type sub dwarfs tabulated 
by Oke \& Gunn (1983) together with Johnson's (1966) colors.

\titleb{Summary}
For the first time an attempt was made to compile all the
currently available observational data on optically 
identified QSO absorber systems. Our main goals are:

\item{$\bullet$} 
We would like to establish 
the status quo of absorber galaxy data
as a basis for further completion of the sample.

\item{$\bullet$} 
We want to demonstrate the usefulness of this sample
for the comparison with evolutionary models for galaxies.
In particular, we study the Hubble diagram for our
models together with observational data from this sample.

\smallskip\noindent
Some important properties of a sample of optically 
identified QSO absorber galaxies should be repeated here:

\item{$\circ$}
Because selection is done with respect to gas absorption cross 
section (cf. Steidel, 1993) this sample is almost unbiased 
with respect to luminosity, surface brightness, radio emission, etc.

\item{$\circ$}
The search for QSO absorbers provides optical data on a large number
of normal galaxies at high redshift. Usually the number of galaxies
found is higher than the number of absorbers searched for, i.e.
neighboring objects or galaxies at somewhat lower or higher
redshift are found in the same field. Their observed properties
are almost as valuable as the ``true'' absorbers for the study 
of evolutionary effects in galaxies and for possible constraints 
on cosmological parameters.

\smallskip\noindent
However, the compilation of our sample from different sources
is connected with some problems:  

\item{$\star$} 
Inhomogeneity of data, e.g. various authors give
apparent magnitudes in different passbands
and use different photometric systems.

\item{$\star$}
Detailed information on the photometric system (filters and
calibrations) is often very difficult to retrieve.
 
\item{$\star$}
From most sources apparent magnitudes are available
in one passband only, i.e. no color information is available. 

\smallskip\noindent
We hope that this pilot study will help continue observations 
to not only identify absorbers but also provide as much
information (colors) as possible for a more detailed
comparison with galaxy evolution models.

\begfigwid 15.0cm
\vskip -18.2cm
\hskip 0.8cm
\vskip 0.3cm 
\figure{2}
{Comparison of observational data with models
for \H0{50}, $\Omega = 1$ and \zform\ $= 5$.
Data points from various publications are distinguished by different
symbols. {\bf Filled} symbols indicate 
{\it spectroscopically confirmed} redshift data whereas 
{\bf open} symbols mark {\it less reliable} absorber {\it candidates}.
Spectroscopically measured redshifts for non absorber
galaxies are marked with bold open symbols.
We emphasize that the distributions of filled and open symbols
are rather similar indicating no systematic deviations of
data for confirmed and candidate absorbers.
Results from galaxy evolution calculations are presented as 
curves with different line--type for respective spectral 
types (indicated as E . . . Sd at the curves).
Luminosity ranges ($2\sigma$ bars), i.e. deviations from 
the model curves, for types E, Sa and Sd are given in the 
lower right corner as in Figure~1
} 
\endfig

\titlea{Comparison with models}
In this Section we compare the data from published observations 
described in \S 3 with the results from our galaxy evolution models 
described in \S 2 using a standard set of cosmological parameters 
(\H0{50}, $\Omega_0 = 1$ and $\Lambda_0 = 0$).

Because of the inhomogeneity and scarcity of published 
data (as discussed in \S 3) we restrict our 
comparison with galaxy evolution models 
to apparent magnitudes R(z), r(z) and g(z) and colors (g--r)(z). 

\titleb{General presentation and results}

In Figure~2 we present data on apparent red magnitudes in the 
Johnson R band as a function of redshift $z$.
Data in the Thuan \& Gunn r band have been transformed 
according to $R = r - 0.4$ as discussed in \S 2.2. 
The whole data set presented was taken from 14 different
sources. Data points from each publication are marked with 
different symbols and the numbers in the legend correspond 
to the references listed in Table~1. Data from papers
1), 6) and 17) do not appear in Figure~2. All observations
reported in 1) (Lanzetta \& Bowen 1990) are covered by
later publications, predominantly by Bergeron \& \Boisse\ (1991).
Nelson \& Malkan (1992) -- paper number 6) -- and Steidel \et\ 
(1994b)  -- paper number 17) -- did not publish any r or R magnitudes.

According to our 
two ``reliability classes'' introduced in \S 3.2.,
{\bf filled symbols} correspond to galaxies with 
{\it spectroscopically confirmed redshifts} and 
{\bf open symbols} denote {\it less reliable} 
absorber {\it candidates}.
Some galaxies turned out to be not the absorber
after spectroscopy had been performed. Thus as a third
class of objects we have to consider {\it spectroscopically 
confirmed non--absorber galaxies}. They are marked with 
an ``bold open symbol'' in Figures~2 and~4.

The results from our evolutionary spectral synthesis code 
are presented, too. Figure~2 shows 6 curves for various 
spectroscopic galaxy types E, S0, Sa, Sb, Sc and Sd, 
as characterized by their respective star formation laws 
(see \S 2.1.).

An important result immediately seen from Figure~2 is that 
virtually all observational data points are lying between 
the curves for E-- and Sd--models. Taking into account the 
luminosity ranges of the various galaxy types 
($2 \sigma_R$ bars in the lower right corner of Figure~2)
as derived from their respective luminosity functions 
we can state that all data 
are bordered by our model curves for E-- and Sd--galaxies 
and accordingly we can establish global conformity between 
our galaxy evolution models and observational data. 

Apparent g--magnitudes g(z) from two references along with 
the respective model calculations for the same standard set 
of cosmological parameters are presented in \fig3a.
This representation of the Hubble diagram confirms 
our results for r-- and R--magnitudes.
Nearly all g--magnitudes lie in the region confined by our
E-- and Sd--curves, however, the number of data points is much
smaller than in Figure~2.
Some absorber {\it candidates} from 
Nelson \& Malkan (1992) (cf. reference 6) in Table~1) 
marked as open circles in \fig3a seem to be brighter than our
E--curve even when the intrinsic luminosity scatter
(indicated by the bar at the E--curve) is taken into account.
However, redshifts of these objects are not measured
spectroscopically but inferred from the assumption 
that they are at the redshift of the QSO absorption system.
If spectroscopy would not confirm the high redshifts
of these objects but instead place them at lower redshift,
they might come to lie within the range of our models.
We note that all $g(z)$ data with {\it spectroscopically 
measured} redshifts are instead bracketed by our models.

Observational data on apparent magnitudes in other 
passbands (e.g. i, V) yield similar results but
are not discussed any further here because of the small
number of data points available.

In \fig3b (g--r) color evolution is presented with data from 
the same references as in \fig3a because they provide the most
comprehensive set of color data available so far (cf. Table~1). 
In both figures numbers are assigned to the data points to identify 
each individual absorber galaxy for further discussion.

(g--r) colors of all our model galaxies (E through Sd) get redder
by about the same amount towards higher redshift until $z \sim 0.5$
and bluer thereafter. The reason for this behavior is that
for $z < 0.5$ the cosmological correction $k_\lambda(z)$
dominates the evolutionary correction $e_\lambda(z)$ for
E--galaxies while for $z > 0.5$ the contrary is true.
For the Sd--model with its SF rate being constant in time,
$e_\lambda(z)$ is always very small.
$k_\lambda(z)$ (\ha $k_g$) for the g band
increases by $\sim 0.7$ mag from $z \sim 0$ to $z \sim 0.4$
with $k_r$ ($k_\lambda(z)$ for the r band)
still remaining negligible. This explains
the reddening of Sd galaxies for z going from $0$ to $\sim 0.4$.
Between $z \sim 0.6$ and $0.8$, $k_r$ increases
to $\sim 0.6$ with $k_g$ remaining close to its $z \sim 0.4$ value.
This accounts for the bluing of (g--r) for $z > 0.4$.

\begfigwid 6.0cm
\vskip -19.3cm
\hskip 0.8cm
{\epsffile{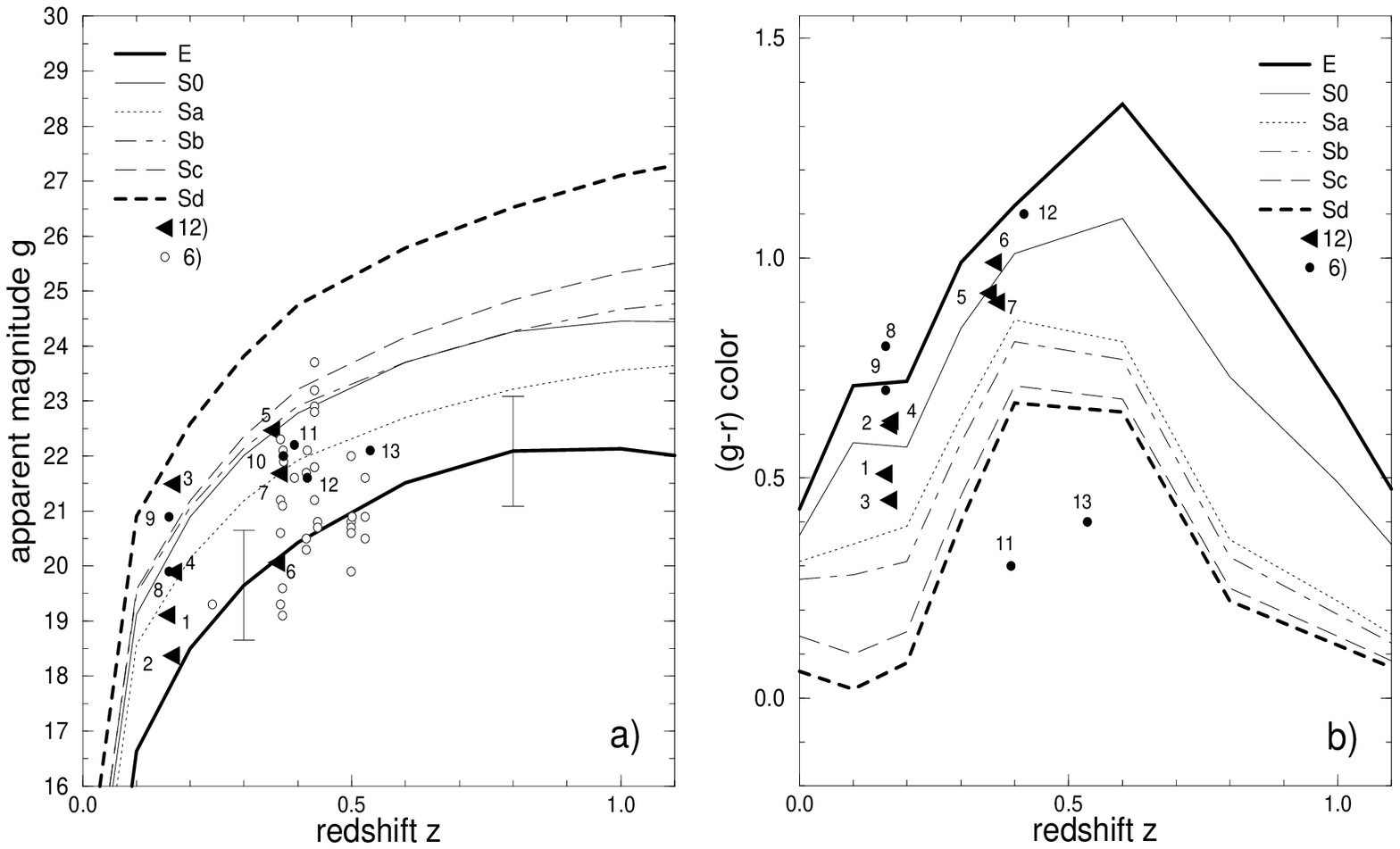}}
\vskip 1.2cm 
\figure{3}
{a) Comparison of observed g(z) data with galaxy evolution models.
Hubble types and references of observations are indicated in the legend.
b) (g--r) colors for the same references and models as in a).
All redshifts for (g--r)(z) data points (references 6) and 12) 
from Table~1.) are spectroscopically confirmed. 
Data points are given the same numbers in both panels
to identify objects for further discussion in the text} 
\endfig
\begfigwid 5.0cm
\vskip -6.0cm
\epsfysize=7.8cm
\hskip 0.2cm
\vskip -7.825cm
\epsfysize=7.8cm
\hskip 5.85cm
\vskip -7.84cm
\epsfysize=7.8cm
\hskip 11.5cm
\figure{4}
{Comparison of observational data with our galaxy evolution 
models for different cosmological parameter 
pairs (h, $\Omega_0$) with \H0{100~h}:
a) (0.5, 0.1), b) (1, 1), and c) (1,0.1). For (0.5, 1) we refer 
to Figure~2. Note that the data points are always the same.} 
\endfig

Colors are more useful than luminosities for the comparison
of models with observations, especially if they cover a long
wavelength range, and thus, in particular, more color data
would be desirable.

\titleb{Evidence for late type absorber galaxies}

Figure~2 suggests that
all spectral types of galaxies from E through Sb and 
possibly all through Sd are present within the 
sample of optically identified absorber galaxies. Most of the
absorber galaxies appear 
to be early through intermediate type spiral types (Sa--Sbc) but 
many ellipticals and late spirals seem to be present, too.
For instance two absorber candidate objects from Yanny \et\ (1990)
(reference 2) in Table~1 and small circles in Figure~2) fall
very close to the Sd--curve.

\fig3a which shows g--magnitudes for the same models 
as Figure~2 reveals one spectroscopically confirmed
object (number 3 in \fig3a) fairly close to the Sd--curve
but the color of this object in \fig3b rather suggests
an early spiral type.

Roughly, from the g(z) -- diagram in \fig3a we would tend 
to attribute an early type to objects 2 and 6,
an intermediate type to objects 4, 5, 7, 8, 10 and 
11 and a late type to object 3.

From the color--redshift diagram (g--r)(z) in \fig3b
we would confirm the early type object classification of objects 2
and 6 and the intermediate type classification of objects 4 and 5.
However, object 8 is too red to be consistent with our 
luminosity -- based type classification, and objects 11 and 13
are much too blue. For object 10 no r--magnitude and
thus no (g--r) color is available.

A more detailed investigation of the individual objects is
required to decide about the nature of those few objects 
which are either too blue or too red to be explained by our
model curves. Our models describe average spectral types of
normal isolated galaxies and are not capable to 
explain these peculiar objects. Additional reddening
might be explained by internal dust 
and the very blue colors could be due to a burst
of star formation perhaps in the course of galaxy 
interaction or the presence of an AGN.

As a word of caution it should be mentioned that even a
galaxy with a redshift \zgal\ $\approx$ \zabs\ cannot
be proven to be the actual absorber, there can always
be a fainter source with still closer projected distance
to the quasar that is not detected. However, as long as
those -- probably very few -- objects in Figure~2 and 3 that
are not the physical absorbers, do not have systematically
different luminosities and colors from the real absorbers,
our conclusions remain valid.

The presence of intermediate and late type galaxies
among QSO absorbers is to some degree unexpected and
has far--reaching consequences.
The first \MgII--absorbers that had been optically
identified by Bergeron \& \Boisse\ (1991) were
$\sim$L$^*$ galaxies with luminosities brighter than those
of intermediate/late type galaxies.

The presence of intermediate and late spectral types among
QSO absorbers, if confirmed by future observations, would
imply that at least an important fraction, if not all, of
these galaxies do have extended gaseous halos with Mg--
and C--column densities high enough to cause absorption
stronger than the lower limit equivalent width for the
detection of an absorption feature. Detailed investigation
of individual objects -- both from a chemical and from
a spectroscopical point of view -- promise interesting
results for masses and extensions of halos (as derived from
impact parameters), dark matter content of galaxies and 
chemical evolution scenarios.
These objects will be an exciting target for HST imaging
as well as a challenge for chemical and dynamical evolution models.

Unfortunately, almost no information about spectral or morphological
types is as yet available for the optically identified QSO absorbers.
Figure~2 can be understood as to predict -- within the 
uncertainties given by the partly overlapping luminosity
ranges of the different galaxy types -- the spectral 
types of individual absorbers.
These predictions can directly be tested by spectral
observations, which, however, for these faint objects, 
requires longer integration times than those generally
applied to just obtain a redshift.

If the direct relation between spectral and morphological
type observed for nearby galaxies would extend to high
redshifts, then Figure~2 could be used to predict 
morphological types for individual absorbers.
Deep HST imaging should allow to assess this
interesting question for redshifts up to $z \sim 0.5$.

Interestingly, Dickinson (1995) reports recent HST
observations of the $z = 0.44$ absorber in front
of 4C06.41 which directly reveal a spiral morphology.
This observation is part of a large optical 
identification project of \MgII\ absorbers which 
may reveal more such exciting results in the near future.
Further evidence for extensive dark halos around two
nearby ($z = 0.092$ and $0.075$)
low--luminosity spiral galaxies inferred 
from QSO absorption line studies was reported by 
Barcons \et\ (1995). However, data are 
still scarce and this conclusion 
remains to be confirmed by future observations.

\titleb{Discussion of cosmological parameters}

Our galaxy evolution models enable us to study the influence 
of different parameters ($H_0$, $\Omega_0$) in the standard
Friedmann--\Lemaitre\ cosmology. We consider four different 
parameter combinations (h, $\Omega_0$) of the Hubble constant 
(\H0{100~h}) and the density parameter $\Omega_0$ in the 
Hubble--diagram for red apparent magnitudes R(z). 

(h, $\Omega_0$)\ $ = $\ (0.5, 1) is already shown in Figure~2 and
the parameter combinations (0.5, 0.1), (1, 1) and (1, 0.1)
are presented in Figure~4 panels a) to c).
Only our E--, Sa-- and Sd--models are depicted for clarity.

In the case of \H0{50} and $\Omega_0 = 0.1$ (\fig4a) a large 
number of galaxies are by up to two magnitudes brighter than 
the E--galaxies in our model. In other words, calculations 
with this cosmological model result in galaxies much too faint 
as compared with the observations. Thus, the combination of \H0{50} 
with a low cosmological density parameter $\Omega_0 = 0.1$ 
cannot be reconciled with the data.

Not for ellipticals but for spectral types S0 and later, gaseous
emission and internal extinction do affect observed magnitudes.
These effects are not included in our models but have been
studied by Guiderdoni \& Rocca--Volmerange (1988). They
find that over a range in redshift $z = 0 \ldots 2$ and in
wavelength from V through I nebular emission does at most
brighten a galaxy by 0.1 mag, while internal extinction
can significantly dim early spiral types (e.g. by $\sim 0.5$
mag in V and $\sim 0.4$ mag in R for S0's at $z \sim 1.6$
cf. Guiderdoni \& Rocca--Volmerange (1988), seen at random 
orientation and less for later spectral types and/or different
redshifts). 

\fig4b and c) show similar comparisons in the case of \H0{100}.
In both cases, $\Omega_0 = 1$ and $\Omega_0 = 0.1$, the 
observations fall fairly well within the envelopes given by
the model E-- and Sd--galaxies, i.e. in combination with a high
value for $H_0$ both high and low density parameters seem
to be compatible with the data.
In a later investigation with more data on colors available
we hope to be able to put further constraints on cosmological
parameters.

The difference in luminosity evolution between high and low 
cosmological density which we find in our models for early
type galaxies is larger than the one reported in earlier
work by Guiderdoni \& Rocca--Volmerange (1987) and 
others using Bruzual's models (e.g. Spinrad \& Djorgovski, 1987).
For instance at a redshift of 2, our low $\Omega_0$ E--model
predicts $m_R \simeq 24.3$ while for $\Omega_0 = 1$ we find
$m_R \simeq 21.3$. This large difference is due to the fact,
that we use more recent stellar evolutionary tracks from the
Geneva group. Interestingly, our results agree quite well
with those of Bressan \et\ (1994) who use a recent set of
stellar tracks from the Padova group, reflecting the fact,
that over the last years, these independent approaches of stellar
evolutionary modeling have converged to large extent.
Most important changes have occured for low--mass stars,
explaining why towards later spectral types the differences
to earlier models get much smaller.

The Sd model with its constant SF rate only shows little
evolution, the cosmological corrections, too, are small,
and both largely compensate for each other. Thus, the
magnitude difference between models for $\Omega_0 = 1$
and $\Omega_0 = 0.1$ is dominated by the difference of 
the bolometric distance module for both cases.

\rr{For illustrative purposes we chose to present here 
the extreme values \H0{50} and \H0{100} discussed today.
We have, of course, examined a large variety of combinations
of parameters $H_0$, $\Omega_0$ and \zform. At this
preliminary stage, we only briefly want to state that
for \H0{75} a density parameter as low as $\Omega_0 = 0.1$
seems to be excluded by the data. $\Omega_0 = 0.5$ is
marginally allowed, provided that the absorber at $z\approx 1$
(cf. Figures~2 \& 4) and the candidates at $z\approx 1.67$ and
$z\approx 1.79$ are all from the high luminosity tail of the
elliptical galaxy luminosity function. Density parameters
$\Omega_0 > 0.5$ to $\Omega_0 = 1$ are easily compatible
with all available data.}

We have also studied models with a Salpeter--IMF and
find that this would not alter our conclusions 
concerning cosmological parameters.
It does make galaxies brighter by a constant value
of $\sim 1$ mag both in V and R over all times
from 2 through 16 Gyr without changing the colors.
As galaxies at their present age -- as given by
($H_0$, $\Omega_0$, \zform) -- are normalized to
mean Virgo luminosities this constant brightening drops
out over the whole redshift range from $z = 0 \ldots 2$.
Only if the IMF would be allowed to significantly
change with time, it might change our conclusions.

\titlea{Discussion}

\titleb{Problems with observational data}

The compilation of apparent magnitude and color data 
as a function of redshift for optically identified QSO 
absorber galaxies is rendered more difficult by two main problems.

First, various authors use {\bf different photometric 
systems} and do not always give all the 
{\bf information on filters and calibrations} 
necessary for our attempt to compile data from various
investigators into one large comprehensive data base. This makes it 
difficult to compare data from different publications
with our results from galaxy evolution models.
As a first approach we use rough transformations between
different photometric systems as discussed in \S 3..
Because most of the present data are given in the Johnson or 
Thuan \& Gunn filters we restricted our comparison
to these two photometric systems, but we
stress, that in principle, apparent magnitudes can easily 
be calculated from model spectra for any photometric system 
if the filter functions are known as discussed in \S 2.
This will be done in future projects intended to study 
individual objects in detail where more
accurate data will be necessary. 

The second main problem concerns the {\bf determination of 
redshifts if no spectroscopic data are available}.
Thus in \S 3.2., we divided the redshift data in 
two ``reliability classes'' ({\it spectroscopically 
confirmed} and {\it candidates}).
Objects with spectroscopic redshifts, however, do
seem to show the same distribution in our Hubble diagrams
for R and g as those for which the redshifts were only estimated.
In many cases the absorber galaxy cannot be 
clearly identified, i.e. several ``possible candidates''
are discussed because more than one galaxy is close to 
the QSO sight line. In these cases more than one absorber candidate
is given for some particular redshift and the real physical 
absorber cannot be identified. In these cases we 
include several candidates in Figure~2, 3a) and 4 
(symbols with exactly the same z value) to see 
if any of them lie outside the ranges of our model curves.
It seems improbable that the ``true absorbers'' should 
deviate systematically in luminosity or color from those
galaxies that do not cause the absorption but lie close
to the absorber both in redshift and projected separation.
Another possibility
is that more than one galaxy contributes to the absorption
system, i.e. there is a group or cluster of galaxies.
 
\titleb{Groups and clusters of galaxies}
 
The number of spectroscopically measured galaxy redshifts
does not need to be equal to the number of confirmed 
optically identified QSO absorbers.
On the one hand, the redshifts
\zabs\ and \zgal\ may be different indicating that the respective
galaxy cannot be the absorber (open bold symbols in Figure~2) .
On the other hand, there may be more
than one spectroscopically investigated galaxy with
a redshift \zgal\ $ \approx $ \zabs. The latter
case indicates a group or the outskirts of a cluster
of galaxies, whereas either only one galaxy, 
probably the closest to the line 
of sight, or the halos from the members of a group or
cluster may be responsible for the absorption system.

For instance, in \fig3a objects 1 to 4 from Spinrad \et\ 
1993 (reference 12 in Table~1) nearly have the same 
(spectroscopically measured !) redshift $z \approx 0.17$ and 
thus seem to be members of a small group or part of a cluster. 
However, no answer can be given to the question, which galaxy 
(or galaxies) from this group actually causes the absorption lines. 
This might even affect our tentative identification
of an Sd galaxy absorber (object 3 in \fig3a).

\titlea{Future prospects}
The purpose of this study was to compile a
comprehensive sample of galaxies detected in various attempts
to optically identify QSO absorbers and to compare
them to our galaxy evolution models. In this first 
approach we studied Hubble diagrams for 
apparent magnitudes R and g as well as color vs. redshift 
relations using all galaxies found, i.e. confirmed 
absorbers as well as candidates.

In further studies, we intend to investigate the extension 
and chemical composition
of galaxy halos for selected individual objects in more detail 
together with spectrophotometric properties.
Data on impact parameters and rest--frame
equivalent widths are already available 
for most published objects.

Furthermore our evolutionary spectral synthesis code 
can be extended for comparison with observations
in the K band using new stellar spectral libraries.
Additionally, our models can be calculated for different
metallicities, especially subsolar ones that may be better
suited for late type galaxies at high redshift in an
attempt to consistently understand chemical abundances of the halo
gas as well as the spectrophotometric appearance of the
visible part of the galaxies.

\titlea{Conclusions}
Optically identified QSO absorption systems as well as
non--absorbing galaxies serendipitously detected
in these deep fields provide a nearly unbiased sample of 
field galaxies with high redshifts. A comprehensive set of 
data on those galaxies has been compiled from the literature 
and is compared with results of our galaxy evolution models.

\item{$\bullet$}
This is the first attempt to put together a
compilation of all observational data 
currently available on optically 
identified QSO absorption systems.
Care is taken to compare observations made in
different filter systems to establish a basis for future 
applications and further completion of the sample.

\item{$\bullet$}
For a standard cosmological model 
(${\rm H}_0 = 50 ~{\rm km~s}^{-1}$ ${\rm Mpc}^{-1}$, 
$\Omega_0 = 1$, $\Lambda_0 = 0$) and a redshift 
of galaxy formation \zform\ $= 5$,
we find good agreement between our galaxy evolution 
models and observed apparent magnitudes  
and colors over a large redshift range ($0 \le z \le 2$).

\item{$\bullet$}
All spectroscopic galaxy types seem to be present 
in the sample of absorber galaxies with the bulk
of the absorbers being of early and intermediate spiral types.
However, even late type spirals seem to cause absorption 
lines in QSO spectra. The correspondence between
spectral and morphological galaxy types at high redshift
remains to be investigated, though first observations at
$z \sim 0.5$ seem to favor it.

\item{$\bullet$}
If confirmed, the fact that intermediate and late type
galaxies also cause absorption in QSO spectra has
far--reaching consequences for the sizes and chemical
enrichment of their halos and for the dark matter content.

\item{$\bullet$}
A combination of cosmological parameters \H0{50} and 
$\Omega_0 = 0.1$ can be excluded from a comparison of 
the data with our models. For a Hubble constant \H0{100},
both high and low density parameters $\Omega_0 = 1$ and
$\Omega_0 = 0.1$ might be admitted by the data.

\item{$\bullet$}
Colors of optically identified absorber galaxies are very 
scarce and, thus, do not allow for any further conclusions.
{\bf More data in two or more passbands} are needed for
further detailed studies of cosmological models. 

\smallskip\noindent
\rr{We caution again that the results presented here depend on
our adopted models and on the input physics used. Both
might be better constrained in the near future.}

\acknow{This work was supported by Verbundforschung
Astronomie/Astrophysik through BMFT grant FKZ~50~0R~90045
and by the DFG through grant Fr316--1.
We thank the anonymous referee and Prof. Dr. J. Lequeux 
for their suggestions that helped to improve the 
presentation of our results.}

\appendix{: \ Notes on individual publications}
In the following  we briefly comment 
on each paper from which we have compiled data.
Numbers of sub-subsections ( 1{\small )} ... 17{\small )} ) 
correspond to the identification number (id. no.) in 
Table~1 and Figures 2 -- 4 as well. 
In particular we discuss some points which are 
important to our purposes.

\bigskip\app{1) Lanzetta \& Bowen 1990 }
\entitle
{Intermediate redshift galaxy halos: Results from QSO absorption lines}
{K.M. Lanzetta and D. Bowen} 
list optically identified \MgII\ absorption systems in their 
TABLE~1 for the purpose to study correlations between equivalent 
widths of \MgII\ absorption lines and impact parameters.
We assume that TABLE~1 in this paper contains all data on 
optically identified (\MgII) absorption systems available 
up to this time and thus we take it as the starting point 
for our search for further data in the literature until now. 
All $m_r(z)$ data listed in TABLE~1 are reported also in later 
papers and we took them from these more topical sources.

\bigskip\app{2) Yanny \et\ 1990 }
\entitle
{Emission--line objects near QSO absorbers. 
 I. Narrow--band imaging and candidate list} 
{B. Yanny, D.G. York and T.B. Williams} 
report on a search for emission--line objects (predominantly 
from \OII) in the environment of 8 QSOs with 
previously known associated \MgII\ 
absorber systems using narrow--band imaging. Data for 37 objects 
are given but in 14 out of the 37 cases only lower limits for R 
magnitudes could be derived from the detection limit.

In a subsequent study (Yanny 1990)
8 of the 37 absorber candidates in 4 QSO fields have been 
investigated using long--slit spectroscopy confirming their 
incidence with the absorber redshifts 
($0.4 \le z_{\rm abs} \le 0.7$). It turns out that
all 8 spectroscopically studied objects are absorbers,
i.e. \zgal\ $=$ \zabs.

\bigskip\app{3) Bergeron \& \Boisse\ 1991}
\entitle
{A sample of galaxies giving rise to \MgII\ quasar absorption systems}
{J. Bergeron and P. \Boisse\ } 
present imaging and spectroscopic observations aimed at
detecting intervening galaxies responsible for low redshift
($z \sim 0.5$) \MgII\ absorption line systems in quasar spectra.
This paper represents a pioneering study on optical identifications 
of absorber systems confirming the intervening galaxy hypothesis. 
14 confirmed identifications, 4 candidates and 3 ``negative 
cases'', i.e. galaxies closest to the QSO line of sight but 
without any absorption system at galaxy redshift are presented. 
The authors give luminosities of galaxies as well as sizes and 
shapes of the absorbing regions.

We include the 14 spectroscopically confirmed absorber galaxies
and 4 candidates in Figure~2 and 4. The three ``negative cases''
from their Table~6 are included as ``candidates'' as well.

\bigskip\app{4) Bechtold \& Ellingson 1992}
\entitle
{\MgII\ absorption by previously known galaxies at $z \approx 0.5$}
{J. Bechtold and E. Ellingson} 
report on
an investigation of 9 QSO spectra using the Multiple Mirror 
Telescope to search for \MgII\ absorption from 
intervening galaxies. This is the reverse procedure:
galaxies and their redshifts ($0.2 \le z \le 0.7$) are 
known from previous studies of the QSO environment 
and now, the possible absorption systems are searched for.
These data are as useful to us as those from
studies aimed to find intervening galaxies near QSOs
with absorption line systems. 

It should be noted that a considerable fraction of the data comes 
from galaxies in high density environment, i.e. galaxy clusters.
This is reflected by the fact that many redshift values are close 
each other. In the cases where \zgal\ $\ne {\rm z}_{\rm QSO}$ is
not confirmed by spectroscopy a physical influence on the galaxies
from the quasar cannot be excluded.

In their \TABLE 4, the authors give r(z) data for 21 galaxies 
in the environment of 9 quasars. Photometric data are taken 
from Green \& Yee (1984), Yee, Green \& Stockman (1986), and 
unpublished images from Yee. Spectroscopy is from Ellingson, 
Green \& Yee (1991) and Ellingson \& Yee (1992). Only two \MgII\ 
absorption--line systems are identified with corresponding galaxy 
redshifts. Consequently most of the r(z) data from this paper
do not belong to optically identified QSO absorbers
and consequently they are not included to our compilation.

In \TABLE 5 of their paper the authors list previously 
known identified absorber galaxies (most from Bergeron 
\& \Boisse\ 1991) and some new candidates. The latter 
ones are included in our compilation.

\bigskip\app{5) Bergeron \et\ 1992}
\entitle
{Discovery of $z \sim 1$ galaxies causing quasar absorption lines}
{J. Bergeron, S. Christiani and P.A. Shaver}
report on the first identification of intervening galaxies
responsible for four \MgII\ absorption--line systems at 
redshifts of about 1 in two QSO fields. There are three candidates
in each field assumed to be responsible for the \MgII\ doublets 
in the QSO spectra. Five galaxy redshifts have been measured
showing incidence with three of the four absorption line redshifts.
Two galaxies are not at any absorber redshift and one galaxy
without measured redshift is assumed to be at the remaining
absorption redshift. The two confirmed non--absorbers are 
marked with open bold line symbols in Figure~2.

\bigskip\app{6) Nelson \& Malkan 1992 }
\entitle
{A photometric search for \OII\ lines in \MgII\ 
 quasar absorption systems}
{B.O. Nelson and M.A. Malkan} 
report on results obtained by narrow-- and broad--band photometry 
of 336 objects near the line of sight to 22 quasars showing \MgII\ 
absorption lines in their spectra. The attention is focused to \OII\ 
emission--line objects which are claimed to undergo enhanced star 
formation and these galaxies are searched for using narrow--band 
photometry. Gunn g magnitudes for 52 objects are given in their 
\TABLE~4. The corresponding absorber redshift can
be derived from their list of observed QSO fields in \TABLE~1.
Some quasars have not been imaged in the Gunn g bandpass but
e.g in ``Spinrad Red''. These data are not included and 
finally 43 g magnitudes remain in our compilation.

In 6 cases redshifts of identified galaxies had previously been 
measured spectroscopically by other authors (Bergeron \& \Boisse\ 
1991, Bergeron \et\  1988, Christiani 1987, Miller \et\  1987, see 
also references in \TABLE 4 of Nelson \& Malkan). 
In most other cases (without spectroscopy) more than
one candidate is given for each absorber redshift.
They are all included in our compilation. In five
cases we are able to adopt r magnitudes from the papers 
mentioned to derive (g-r) colors. These colors are used in Figure~3b.

\bigskip\app{7) Steidel \& Dickinson 1992}
\entitle
{The unusual field of the quasar 3C 336: Identification
 of three foreground \MgII\ absorbing galaxies}
{C.C. Steidel and M. Dickinson} 
present imaging and spectroscopic observations of the field 
of 3C 336 ($z_{\rm em} = 0.927$) which reveals at least 
3 \MgII\ absorption systems with $z_{\rm abs} < z_{\rm em}$.
Data for 6 galaxies with spectroscopically measured redshifts and 
\R\ ($\lambda = 6930$, $\Delta\lambda = 1500$), 
g ($\lambda = 4900$ and $\Delta\lambda = 700$) and
i ($\lambda = 8000$ and $\Delta\lambda = 1450$) AB magnitudes 
are given in their \TABLE~1. 
3 galaxies seem to be the absorbing galaxies. 3 further
objects with \zgal\ $\ne$ \zabs\ are marked with open
bold line symbols in Figure~2 and 4 as usual.

In this paper Steidel \& Dickinson give 
apparent AB--magnitudes \R$_{\rm AB}$ (6930/1500) 
together with $( g - $\R$ )_{\rm AB}$
and $( $\R$ - i )_{\rm AB}$ colors.
Because we restrict our comparison of observations with 
models to Thuan \& Gunn and Johnson bandpasses as discussed 
in \S 2.2 we cannot take the AB colors from Steidel \& 
Dickinson (1992) into consideration.

\bigskip\app{8) Steidel \& Hamilton 1992 }
\entitle
{Deep imaging of high redshift QSO fields below the Lyman limit. 
 I. The field of Q0000-263 and galaxies at $z = 3.4$}
{C.C. Steidel and D. Hamilton} 
present results for the first field completed in a new survey 
involving very deep imaging of the fields of high redshift
quasars with optically thick Lyman limit absorption systems.
In their \TABLE 1 the authors list $U_n$, 
G and \R\ magnitudes for 33 objects within 30'' of Q0000-263 
($z_{\rm em} = 4.106$). The moderately under--luminous galaxy 
at $z = 0.438$ was previously suggested to be at $z = 3.408$ 
(Turnshek \et\ 1991), but now Steidel \& Hamilton suggest 
that the line previously identified as Lyman $\alpha$ is 
probably \OII. The $z = 0.438$ and \R\ $=22.5$ galaxy
is added to our list and is an interesting case for a late
type absorber.
Another galaxy is assumed to be a candidate 
for the DLA system at $z = 3.39$.

\bigskip\app{9) Yanny \& York 1992}
\entitle
{Emission--line objects near Quasi--stellar object absorbers. 
 III.  Clustering and colors of moderate--redshift \HII\ regions}
{B. Yanny and D.G. York} 
report on three deep narrow--band images corresponding to \OII\ 
at the redshift of known QSO absorption--line systems which show 
evidence for enhanced star formation occurring in groups of giant 
\HII\ regions spread over 100 -- 300 \kpc. This paper extends 
their earlier investigation (Yanny \et\ 1990, cf. 3{\small )}~). 
For the three QSOs under study the authors list all galaxies 
in the field. From these tables we picked up 7 absorber 
candidates as suggested by the authors.

\bigskip\app{10) Drinkwater \et\ 1993}
\entitle
{On the nature of \MgII\ absorption line systems in quasars}
{M.J. Drinkwater, R.L. Webster and P.A. Thomas}
present the results of a large R--band imaging survey
of 71 bright ($m_v < 18$) quasars revealing \MgII\ absorption 
lines in their spectra. For 21 QSO fields possible absorber candidates
are found, however, no galaxy redshifts have been measured.
A statistical approach is used to identify galaxies likely 
to be associated with the absorption systems. 
In \TABLE 5 and 7 of their paper the authors give \zabs\ and R 
magnitudes for possible absorber candidates. For some 
galaxies two possible \zabs\ are given. 

\bigskip\app{11) Le Brun \et\ 1993 }
\entitle
{A deep imaging survey of fields around quasars with $z < 1.2$ 
 \MgII\ absorption systems}
{V. Le Brun, J. Bergeron, P. \Boisse\ and C. Christian}
report on an imaging survey of 12 fields around quasars 
with 17 \MgII\ absorption systems in the r band at the CFH telescope.
They give a list (Table 3. in their paper) of altogether 19
absorber candidates with 6 galaxy redshifts being 
spectroscopically confirmed. It turned out that
two of them are not absorber galaxies (\zgal\ $\ne$ \zabs).
The authors emphasize that some of the candidates 
are rather uncertain.

\bigskip\app{12) Spinrad \et\ 1993}
\entitle
{Hydrogen and metal absorption lines in PKS0405--123
 from the halos of low redshift galaxies}
{H. Spinrad, A.V. Filippenko, H.K.C. Yee, E. Ellingson, J.C. Blades, 
 J.N. Bahcall, B.T. Jannuzi, J. Bechtold and A. Dobrzycki}
report on HST ultraviolet and ground--based optical spectra of 
the bright quasar PKS 0405--123 which has two absorption--line 
systems. Spectroscopically measured redshifts and Gunn g and r 
magnitudes are given for 9 galaxies. One redshift is very uncertain. 
The values of galaxy redshifts are grouped near two different 
absorber redshifts ($z_1 = 0.1670$ and $z_2 = 0.3516$) 
indicating small groups of galaxies or outskirts of clusters
of galaxies.

In the $z_2$ system only Ly$\alpha$ absorption is observed
whereas in the $z_1$ system metal lines (\CIV\ and probably
\MgII) are detected as well.

The authors give r--magnitudes and (g--r) colors (Thuan \& 
Gunn (1976) g and r) in their \TABLE~1 and thus g--magnitudes
can be derived.

\bigskip\app{13) Steidel \et\ 1993}
\entitle
{A dwarf galaxy near the sight line to PKS 0454+0356:
 A fading ``faint blue galaxy''?}
{C.C. Steidel, M. Dickinson and D.V. Bowen} 
report on the discovery of a dwarf galaxy at
$z = 0.072$ which is only 4'' in projection from the line of
sight to the quasar . Fortuitously the spectrum of a second
galaxy was recorded simultaneously from several moderately strong
emission lines revealing $z \approx 0.2$. Here we are not interested
in dwarf galaxies and therefore only data from the latter object
are included in our compilation. By now no appropriate absorption
lines are found in the QSO spectrum. Future observations
(using HST) in search of \MgII\ resonance lines would be very 
meaningful for the study of the gas associated with an intrinsically
faint, star--forming dwarf galaxy.

\bigskip\app{14) Aragon-Salamanca \et\ 1994}
\entitle
{The nature of distant galaxies producing multiple \CIV\ 
 absorption lines in the spectra of high-redshift quasars}
{A. Aragon-Salamanca, R.S. Ellis, J.-M. Schwartzenberg and J. Bergeron}
demonstrate that luminous L$^*$ galaxies up to $z \simeq 2$ 
which cause \CIV\ absorption 
in the spectra of quasars can be detected by deep infrared (K band) imaging. 
 
The authors use a statistical approach to identify the 
absorbers from photometry alone because at these faint 
limits spectroscopy is very difficult. A sample 
of 11 QSOs with clustered \CIV\ absorption line systems 
is investigated. In some cases, several suitable
complexes at different \zabs\ are present within a single 
QSO spectrum. As the redshift of possible absorbers we take 
from their \TABLE 1 the mean values $\langle z_{abs}^{\CIV} 
\rangle$ weighted by the appropriate rest--frame equivalent 
width for each system. 

A subset of the QSO fields has also been imaged in R and I. 
Thus, 19 K magnitudes and 9 $(R-K)$ colors
are given in \TABLE 2 of the paper. From these data we obtain
R magnitudes as a function of redshift for 9 absorber candidates.

\bigskip\app{15) Ellingson \et\ 1994 }
\entitle
{Metal--line absorption at \zabs\ $\sim {\rm z}_{\rm em}$ 
 from associated galaxies} 
{E. Ellingson, H.K.C. Yee, J. Bechtold and A. Dobrzycki} 
report on deep optical imaging and limited spectroscopy of 
10 fields around quasars with $0.15 < {\rm z}_{\rm em} < 0.65$
revealing \CIV\ absorption predominantly at the emission
redshift of the QSO. The aim of their investigation is not
the optical identification of absorber systems but the
study of the interaction of quasars with galaxies from
the surrounding clusters. Some of their data, compiled
in \TABLE 2. are also useful to our purposes. We select 
7 candidates from their list, one galaxy with
spectroscopically confirmed redshift. 

We want to caution that the rich cluster environment and the 
proximity to the quasars might be influential on 
the evolution of these galaxies. 

\bigskip\app{16) Steidel \et\ 1994a}
\entitle
{The z = 0.8596 damped lyman alpha absorbing galaxy toward PKS 0454+0356}
{C.C. Steidel, D.V. Bowen, J.C. Blades and M. Dickinson}
present HST and ground--based data on a new DLA system at 
\zabs$ = 0.8596$ along the line of sight to PKS 0454+0356.
Optical images were obtained as part of a large survey for
galaxies producing \MgII\ absorption line systems, including 
a very deep \R\ (6930/1500) image from which we take the \R\ 
magnitude as a function of redshift for this galaxy. The authors 
only gave a faint limit ($K > 20.5$) for the K magnitude.

\bigskip\app{17) Steidel \et\ 1994b}
\entitle
{Imaging of two damped lyman alpha absorbers at intermediate redshift}
{C.C. Steidel, M. Pettini, M. Dickinson and S.E. Persson}
report on observing the fields of the QSOs 3C 286 and PKS 1229--021
in the $I$ and $K_s$ bands. Under good seeing conditions they
resolve faint galaxy images close to the QSO sight--lines.
We adopt I and K magnitudes as a function of redshift for
three absorber candidates from this paper.

\begref{References}
\ref Aragon--Salamanca, A., Ellis, R.S., Schwartzenberg, J.-M., 
     Bergeron, J., 1994, ApJ 421, 27
\ref Barcons, X., Lanzetta, K.M., Webb, J.K., 1995, Nat 376, 321
\ref Bechtold, J., Ellingson, E., 1992, ApJ 396, 20
\ref Bergeron, B., Boulade, O., Kunth, D., Tytler, D., Boksenberg, a.,
     Vigroux, L., 1988, A\&A 191, 1
\ref Bergeron, J., \Boisse, P., 1991, A\&A 243, 344
\ref Bergeron, J., Cristiani, S., Shaver, P.A., 1992, A\&A 257, 417
\ref Bressan, A., Chiosi, C., Fagotto, F., 1994, ApJS 94, 63
\ref Bruzual, G.A., Charlot, S., 1993, ApJ 405, 538
\ref Cristiani, S., 1987, A\&A 175, L1
\ref Dickinson, Marc, 1995, in ``New Light on galaxy Evolution'', eds.
R. Bender \& R. Davis, IAU Symp. No. 171, in press
\ref Drinkwater, M.J., Webster, R.L., Thomas, P.A., 1993, AJ 106, 848
\ref Ellingson, E., Yee, H.K.C., Bechtold, J., 
     Dobrzycki, A., 1994, AJ 107, 1219
\ref Fritze -- von Alvensleben, U., 1989, PhD Thesis, University of G\"ottingen
\ref Fritze -- von Alvensleben, U., Kr\"uger, H.,
Fricke, K.J., Loose, H.-H., 1989, A\&A 224, L1
\ref Fritze -- von Alvensleben, U., Kr\"uger, H.,
Fricke, K.J., 1991, A\&A 246, L59
\ref Fritze -- von Alvensleben, U., Gerhard, O.E., 1994, A\&A 285, 751
\ref Fritze -- von Alvensleben, 1994, in: 
``Stellar populations'', eds. P.C. van der Kruit \& G. Gilmore,
IAU Symp. No. 164, Dordrecht 1994
\ref Guiderdoni, B., 1986, PhD Thesis, Universit\'e Paris VII.
\ref Guiderdoni, B., Rocca--Volmerange, B., 1987, A\&A 186, 1
\ref Guiderdoni, B., Rocca--Volmerange, B., 1988, A\&A Suppl. Ser. 74, 185
\ref Gunn, J.E., Stryker, L.L., 1983, ApJ Suppl. 52, 121
\ref Johnson, H.L., 1966, Ann. Rev. Astron. Astrophys., 4, 193
\ref Kennicutt, R.C., 1992, ApJS 79, 255
\ref Kr\"uger, H., Fritze -- von Alvensleben, U., Loose, H.--H., 
Fricke, K.J., 1991, A\&A 242, 343
\ref Kr\"uger, Fritze -- von Alvensleben, U., Loose, H.--H., 1995, A\&A 303, 41
\ref Lanzetta, K.M., Bowen, D., 1990, ApJ 357, 321
\ref Le Brun, V., Bergeron, J., \Boisse, P., Cristiani, C., 1993, A\&A 279, 33
\ref Lilly, S.J., Cowie, L.L., Gardener, J.P., 1991, ApJ 369, 79
\ref Maeder, A., 1991, A\&A 242, 93
\ref Miller, J.S., Goodrich, R.W., Stephens, S.A., 1987, AJ 94, 633
\ref Nelson, B.O., Malkan, M.A., 1992, ApJS 82, 447 
\ref Oke, J.B., Gunn, J.E., 1983, ApJ 266, 713
\ref Rocca--Volmerange, B., Guiderdoni, B., 1988, A\&A Suppl. Ser. 75, 93 
\ref Sandage, A., Binggeli, B., Tammann, G.A., 1985a, AJ 90, 395
\ref Sandage, A., Binggeli, B., Tammann, G.A., 1985b, AJ 90, 1759
\ref Sandage, A., 1986, A\&A 161, 89
\ref Scalo, J.M., 1986, Fundam. Cosmic Phys. 11, 1
\ref Spinrad, H., Djorgovski, S., 1987, in ``Observational Cosmology'',
eds. A. Hewett, G. Burbidge \& L.Z. Fang, IAU Symp. No. 124, Dordrecht 1987
\ref Spinrad, H., Filippenko, A.V., Yee, H.K.C., Ellingson, E.,
     Blades, J.C. Bahcall, J.N., Jannuzi, B.T., Bechtold, J.,
     Dobrzycki, A., 1993, AJ 106, 1
\ref Steidel, C.C., Dickinson, M., 1992, ApJ 394, 81
\ref Steidel, C.C., Hamilton, D., 1992, AJ 104, 941
\ref Steidel, C.C., 1993, in: Shull, J.M. and Thronson, H.A. (eds.):
     The Environment and Evolution of Galaxies, 263
\ref Steidel, C.C., Dickinson, M., Bowen, D.V., 1993, 
     ApJ Lett. 413, L77
\ref Steidel, C.C., Hamilton, D., 1993, AJ 105, 2017
\ref Steidel, C.C., Bowen, D.V., Blades, J.C., Dickinson, M., 1994a, 
     MIT Preprint Series No. CSR--94-24
\ref Steidel, C.C., Pettini, M., Dickinson, M., Persson, S.E., 1994b, 
     MIT Preprint Series No. CSR--94-24
\ref Steidel, C.C., 1995, in: G. Meylan (ed.), QSO absorption lines,
Proceedings of the ESO workshop on QSO absorption lines held at
Garching, Germany, 21. -- 24. November 1994, Berlin/Heidelberg 1995
\ref Thuan, T.X., Gunn, J.E., 1976, PASP 88, 543
\ref Turnshek, D.A., Machetto, F., Bencke, M.V. , Hazard, C.,
     Sparks, W.B., McMahon, R.G., 1991, ApJ 382, 26
\ref Wu \et\ (eds.) 1983, IUE Ultraviolet Spectral Atlas, {\it NASA} No. 22
\ref Yanny, B., 1990, ApJ 351, 396
\ref Yanny, B., York, D.G., Williams, T.B., 1990, ApJ 351, 377
\ref Yanny, B., York, D.G., 1992, ApJ 391, 569
\endref
\bye